\documentclass[11pt,danish,a4paper,floatfix]{article}
\usepackage{jcappub}

\usepackage{bm}
\usepackage{amsfonts}
\usepackage{subfigure}

\newcommand{\be}{\begin{equation}}
\newcommand{\ee}{\end{equation}}
\newcommand{\bea}{\begin{eqnarray}}
\newcommand{\eea}{\end{eqnarray}}

\renewcommand{\vec}[1]{\boldsymbol{#1}}

\usepackage{lmodern}
\usepackage{slantsc}
\newcommand{\CONCEPT}{\textsc{co\textsl{n}cept}}
\newcommand{\CLASS}{\textsc{class}}
\newcommand{\GADGET}{\textsc{gadget}}
\newcommand{\PKDGRAV}{\textsc{pkdgrav}}
\newcommand{\RAMSES}{\textsc{ramses}}
\newcommand{\COSIRA}{\textsc{cosira}}

\begin{document}


\title{Fully relativistic treatment of light neutrinos in \bm{$N$}-body simulations}

\author[a]{Thomas Tram,}
\author[b]{Jacob Brandbyge,}
\author[b]{Jeppe Dakin,}
\author[b]{Steen Hannestad}

\affiliation[a]{Aarhus Institute of Advanced Studies (AIAS), Aarhus University, DK--8000 Aarhus C, Denmark}
\affiliation[b]{Department of Physics and Astronomy, Aarhus University,
 DK-8000 Aarhus C, Denmark}

\emailAdd{thomas.tram@aias.au.dk}
\emailAdd{jacobb@phys.au.dk}
\emailAdd{dakin@phys.au.dk}
\emailAdd{sth@phys.au.dk}

\abstract{
Cosmological $N$-body simulations are typically purely run with particles using Newtonian equations of motion. However, such simulations can be made fully consistent with general relativity using a well-defined prescription.
Here, we extend the formalism previously developed for $\Lambda$CDM cosmologies with massless neutrinos to include the effects of massive, but light neutrinos. We have implemented the method in two different $N$-body codes, \CONCEPT{} and \PKDGRAV{}, and demonstrate that they produce consistent results. We furthermore show that we can recover all appropriate limits, including the full GR solution in linear perturbation theory at the per mille level of precision.
}

\maketitle


\section{Introduction}

In the coming few years, new, large galaxy surveys such as those from LSST \cite{LSST} and EUCLID \cite{EUCLID} will provide extremely precise measurements of the large scale structure of our Universe.
This in turn requires numerical simulations of structure formation to be accurate at the sub-percent level over a wide range of scales.

One important ingredient in this quest is to include massive neutrinos which are known to make up at least $0.1\,\%$ of the total energy density at present. Even at this lower limit the inclusion of neutrinos changes the matter power spectrum at the $3$--$4\,\%$ level, substantially more than the required precision of these surveys.

Over the past decade, a substantial effort has been devoted to the inclusion of massive neutrinos in $N$-body simulations.
One approach is to use a particle representation of the full neutrino distribution function (e.g.\ \cite{Brandbyge:2008rv,Viel:2010bn,Agarwal:2010mt,Bird:2011rb,Villaescusa-Navarro:2013pva,Castorina:2015bma,Emberson:2016ecv,Adamek:2017uiq,Banerjee:2018bxy,Brandbyge:2018tvk}). This, however, is very numerically challenging because of the large number of particles needed to properly follow the neutrino distribution function.
Another scheme assumes that neutrino perturbations remain linear~\cite{Brandbyge:2008js,AliHaimoud:2012vj,Liu:2017now}. A simple scheme which is known to work well for small neutrino masses is to use the linear neutrino density field calculated by realising the linear neutrino transfer function on a grid \cite{Brandbyge:2008js}. An improvement on this is to solve the linear theory neutrino equations, but use the full non-linear gravitational potential calculated in the simulation~\cite{AliHaimoud:2012vj,Liu:2017now}.
However, in both cases this scheme only works for relatively small neutrino masses where neutrino perturbations remain linear at all times. 
Finally, there are hybrid schemes coupling the two approaches 
\cite{Brandbyge:2009ce}, as well as approaches based on other approximate solutions (e.g.\ \cite{Banerjee:2016zaa,Dakin:2017idt})

Another effect which must be taken into account comes from the inclusion of general relativistic effects. This can be done fully relativistically in the weak field limit (see e.g.\ \cite{Adamek:2016zes}). However, as has been shown (see e.g.\ \cite{Fidler:2017pnb} and references therein), even $N$-body codes such as \GADGET{} \cite{Springel:2005mi}, \PKDGRAV{} \cite{Potter:2016ttn} and \RAMSES{} \cite{Teyssier:2001cp} which are inherently Newtonian, can in fact be used to obtain results which are valid in the weak field limit of GR. In the case of pure $\Lambda$CDM models, i.e.\ models with only one matter component, this can be done via backscaling. The inclusion of massive neutrinos complicates matters, and the backscaling method becomes highly non-trivial.
However, massive neutrinos can be included using the method presented in \cite{Fidler:2015npa,Fidler:2016tir,Brandbyge:2016raj}. This requires neutrinos to be light enough that they can be treated as a purely linear component. In this case the neutrino density field can be realised at each timestep in a Newtonian simulation and thus be used to calculate the neutrino contribution to the local gravitational potential. This method was first introduced in 
 \cite{Brandbyge:2008js} and shown to lead to sub-percent errors in the calculation of the matter power spectrum for neutrino masses up to around $0.3\,\mathrm{eV}$.
When neutrinos are added to the simulation using this method the $N$-body simulation still contains only one matter component and this makes it possible to use the framework presented in \cite{Fidler:2015npa,Fidler:2016tir,Brandbyge:2016raj}.

In this paper we show that by extending the method to include massive neutrinos we can run Newtonian $N$-body simulations which are fully consistent with GR, including massive neutrinos, without compromising the speed and scalability of standard $N$-body codes.
We test our framework using two different $N$-body codes and demonstrate that we obtain fully consistent results.

In Section 2 we discuss the theoretical set-up needed to include massive neutrinos and GR. In Section 3 we present our numerical results, and finally Section 4 contains a discussion and our conclusions.

\section{Method and implementation}

As was shown in \cite{Fidler:2015npa,Fidler:2016tir,Brandbyge:2016raj}, Newtonian $N$-body simulations containing only dark matter (or any other highly non-relativistic component) can be made compatible with general relativity. 

For pure matter (i.e.\ a pressureless component) the continuity and Euler equations for the density contrast $\delta$ and peculiar velocity $\vec{v}$ can be written as 
\begin{eqnarray}
\dot\delta + \nabla \cdot {\vec v} & = & 0\,, \\
( \partial_\tau + {\cal H} ) {\vec v} & = & -\nabla \phi + \nabla \gamma\,,
\end{eqnarray}
where a dot denotes differentiation with respect to conformal time $\tau$ and ${\cal H} = \dot{a}/a$ is the conformal Hubble parameter with $a$ being the cosmic scale factor. The quantity $\gamma$ is a correction which can be subtracted from the peculiar potential, $\phi$, in the simulation. The potential $\phi$ is the total potential from all species, i.e.\
\begin{equation}
	\nabla^2\phi = \nabla^2\sum_{\alpha} \phi_\alpha = 4\pi G a^2 \sum_\alpha \delta\rho_\alpha\,,
\end{equation}
with $\alpha\in\{\text{cdm}, \text{b}, \gamma, \nu\}$ running over all species\footnote{The subscript $\gamma$ refers to photons and should not be confused with the variable $\gamma$, representing the relativistic potential correction.}.

From \cite{Fidler:2017pnb}, the Fourier space equation for $\gamma$ can be written as
\begin{equation}
\gamma k^2  = -(\partial_\tau + {\cal H} ) \dot H_\text{T} + 8 \pi G a^2 \Sigma\,, \label{eq:gamma_from_perturbations}
\end{equation}
where $\Sigma$ is the total anisotropic stress of all species and $H_\text{T}$ is the trace-free component of the spatial part of the metric in $N$-body gauge
(see e.g.\ \cite{Adamek:2017grt}).
In appendix~\ref{appendix:gamma} we calculate $\gamma$ in Fourier space with massive neutrinos included. We then have everything we need for this approach to fully consistently take massive neutrinos into account.

Concretely we split the total potential $\phi - \gamma$ experienced by the matter in the simulation into a contribution coming from the matter itself (calculable using standard techniques in the $N$-body simulation), $\phi_{\text{sim}}$, and a contribution coming from photons, neutrinos, and the GR correction $\gamma$, $\phi_{\text{GR}}$:
\begin{equation}
	\phi - \gamma \equiv \phi_{\text{sim}} + \phi_{\text{GR}}\,,
\end{equation}
with $\phi_{\text{GR}}$ given by
\begin{align}
	\nabla^2\phi_{\text{GR}} &\equiv \nabla^2\bigl( \phi_\gamma + \phi_\nu - \gamma \bigr) \notag \\
	&\equiv 4 \pi G a^2 \bigl( \delta\rho_{\gamma} + \delta\rho_{\nu} + \delta\rho_{\text{metric}} \bigr) \label{eq:Poissson_GR} \\
	&\equiv 4\pi G a^2 \delta\rho_{\text{GR}}\,. \notag
\end{align}
Here $\delta\rho_{\text{metric}}$ is a fictitious density perturbation which amounts to the GR potential correction $\gamma$,
\begin{equation}
	\nabla^2 \gamma = -4\pi Ga^2 \delta\rho_{\text{metric}}\,. \label{eq:gamma_Poisson}
\end{equation}
Following the same prescription as in \cite{Brandbyge:2016raj},
at each timestep in the simulation we realise $\delta\rho_{\text{GR}}$ in Fourier space, solve its Poisson equation \eqref{eq:Poissson_GR}, transform to real space and apply the force from $\phi_{\text{GR}}$ to the matter particles, in addition to the usual force from the matter particles themselves (corresponding to $\phi_\text{sim}$).

To compute $\delta\rho_{\text{GR}}$ in linear perturbation theory, a \CLASS{} \cite{Blas:2011rf} computation has been run in advance, providing us with $\delta \rho_\gamma$ and $\delta \rho_\nu$ in either synchronous or conformal Newtonian gauge, $\dot{H}_{\text{T}}$ in $N$-body gauge as described in appendix~\ref{appendix:gamma}, as well as $\Sigma$ (which is gauge independent in linear perturbation theory and can be calculated from its algebraic relation to $\phi-\psi$ in conformal Newtonian gauge), all as functions of $a$ and $k$. From $\dot{H}_{\text{T}}^{\text{Nb}}$ and $\Sigma$, we obtain $\delta\rho_{\text{metric}}^{\text{Nb}}(a, k)$ using \eqref{eq:gamma_from_perturbations} and \eqref{eq:gamma_Poisson}. We then transform $\delta\rho_\gamma$ and $\delta\rho_\nu$ to $N$-body gauge,
\begin{equation}
	\delta\rho_\alpha^{\text{Nb}} = \delta\rho_\alpha^{\text{S/N}} + 3{\cal H}(1+w_\alpha)\frac{\theta_{\text{tot}}^{\text{S/N}}}{k^2}\bar{\rho}_\alpha\,,
\end{equation}
with $\theta_{\text{tot}}$ the total peculiar velocity divergence of all species and $w_\alpha=\bar{P}_\alpha/\bar{\rho}_\alpha$ the equation of state parameter of species $\alpha$ (both obtainable from \CLASS{}), after which we add $\delta\rho_\gamma^{\text{Nb}}$ and $\delta\rho_\nu^{\text{Nb}}$ to $\delta\rho_{\text{metric}}^{\text{Nb}}$, resulting in $\delta\rho_{\text{GR}}^{\text{Nb}}$. The realisation of this $\delta\rho_{\text{GR}}^{\text{Nb}}(k)$ on a grid in real space is done using the formalism outlined in appendix A of \cite{Dakin:2017idt}.

\begin{figure}[t]
\begin{center}
\includegraphics[width=0.8\textwidth]{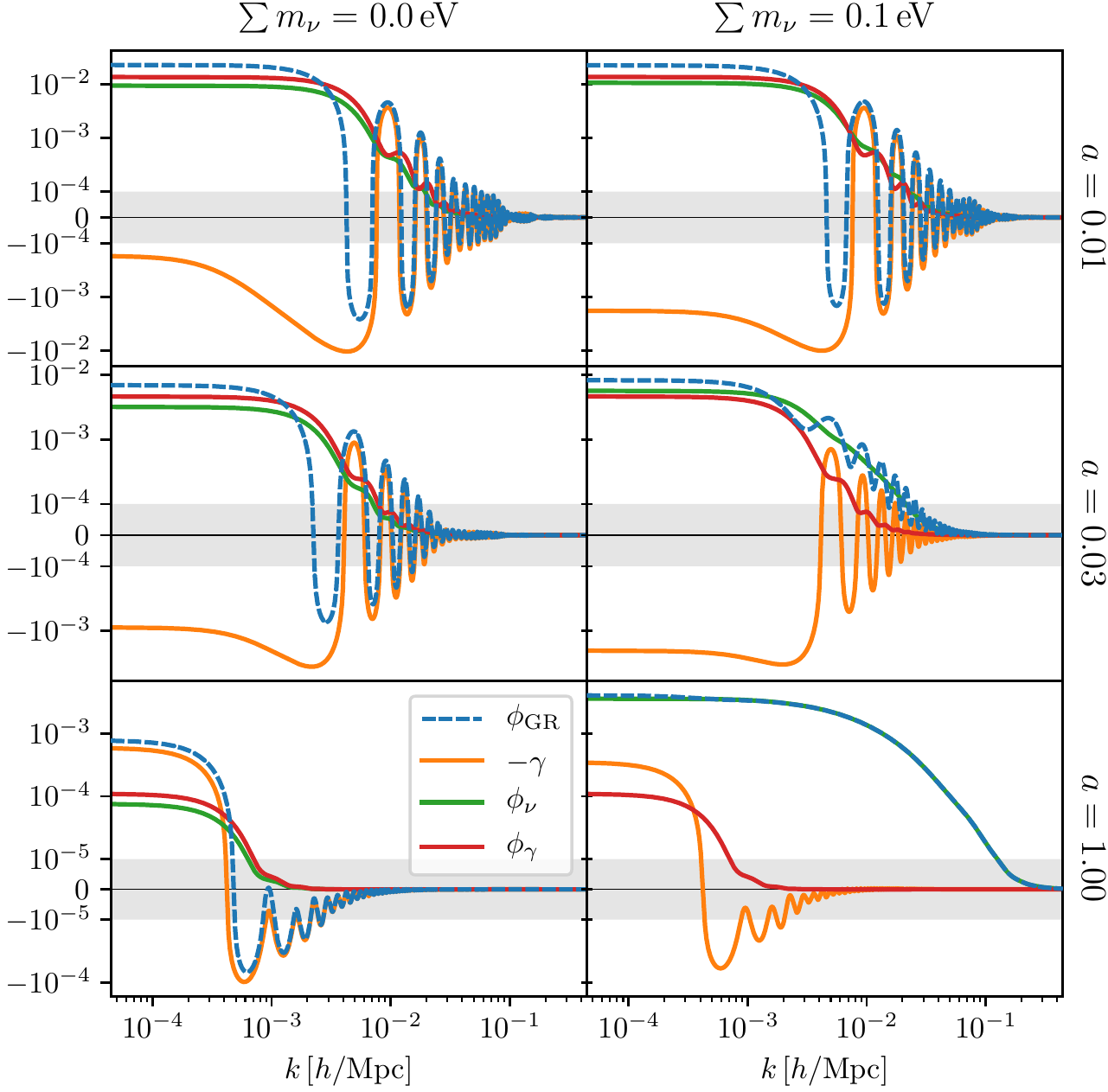}
\end{center}
\caption{Individual contributions to $\phi_{\rm GR}\equiv \phi_\gamma + \phi_\nu -\gamma$ at three different scale factors. The left plot shows the case of massless neutrinos, the right plot shows the case of $\sum m_\nu=0.1\,\mathrm{eV}$. The potentials are all in $N$-body gauge. The gray bands indicate regions where the vertical axes scale linearly.
\label{fig:gamma}}
\end{figure}

In Fig.~\ref{fig:gamma} we show the individual contributions to $\phi_{\rm GR}$ from photons, neutrinos, and the GR correction $\gamma$.
For the case of massless neutrinos we reproduce the results from 
\cite{Brandbyge:2016raj}: For small (superhorizon) values of $k$ all three contributions asymptotically approach $k$-independent values, while for larger $k$ all three contributions oscillate and damp.
For the case of massive neutrinos, we see that, as expected, the neutrino contribution ceases to oscillate as soon as neutrinos become non-relativistic 
($T/m_\nu \sim 1/3$ around $z \sim 60$). From this point on it grows rapidly, essentially following the matter evolution.
We also note that the photon and $\gamma$ contributions remain almost unchanged in the case of massive neutrinos so that by far the largest difference between models with different neutrino masses comes from the neutrino component itself, rather than from photons or the GR correction $\gamma$.

We initialise the simulation using \CLASS{} in the same way as was described in \cite{Dakin:2017idt}. Initial conditions for the density and velocity fields are generated using the $N$-body gauge $\delta_{\text{cdm}+\text{b}}$ and $\theta_{\text{cdm}+\text{b}}$ transfer functions from \CLASS{} at the initialisation time, i.e.\ we do not use higher order methods such as 2LPT. In this particular case this is completely unproblematic since we study effects pertaining to very large scales where structures are completely linear at the initial time.

\section{Numerical setup and results}

In order to test the effect of massive neutrinos including GR corrections we perform a suite of $N$-body simulations, primarily using the publicly available \CONCEPT{} $N$-body solver \cite{Dakin:2017idt}. All \CONCEPT{} simulations in this work use cosmological parameters as listed in table~\ref{table:class_parameters}. We use a degenerate neutrino hierarchy, i.e.\ three neutrinos of equal mass. The \CONCEPT{} simulations all begin at $a=0.01$, use $1024^3$ matter particles and the potential grids (both $\phi_{\text{sim}}$ and $\phi_{\text{GR}}$) are of size $1024^3$. All \CONCEPT{} simulations are carried out in box sizes of either $(16384\,\text{Mpc}/h)^3$ or $(1024\,\text{Mpc}/h)^3$, the power spectra from which are patched together to give the ones shown in Fig.~\ref{fig:relpower1}, \ref{fig:relpower2} and \ref{fig:relpower3}.

\begin{table}[tb]
    \begin{center} 
        \begin{tabular}{l c c} 
            \hline
            Parameter & $\Lambda$CDM  & $\sum m_\nu = 0.10\,\text{eV}$  \\
            \hline
            $A_\text{s}$  & $2.215 \times 10^{-9}$& $2.215 \times 10^{-9}$ \\
            $n_\text{s}$ & $0.9655$ & $0.9655$ \\
            $\tau_\text{reio}$ & $0.0925$ & $0.0925$  \\
            $\Omega_{\text{b}}$ & $0.049$  & $0.049$  \\
            $\Omega_{\text{cdm}}$ & $0.264$  & $0.262$ \\
            $\Omega_{\nu}$ & $3.77\times 10^{-5}$ & $2.37\times 10^{-3}$  \\  
            $h$ & $0.6731$ & $0.6731$ \\
            \hline
            $N_{q,\nu}$ & $1000$ & $1000$ \\  
            $l_{{\rm max},\nu}$ & $1000$ & $1000$ \\  
            \hline						
        \end{tabular}
    \end{center}
    \caption{Cosmological parameters and numerical settings for the \CLASS{} runs used. We use the exact relation $\Omega_{\text{cdm}} = 0.2643 - \Omega_{\nu}$.}
    \label{table:class_parameters} 
\end{table}

\subsection{Main results}

\begin{figure}[t]
\begin{center}
\includegraphics[width=0.8\textwidth]{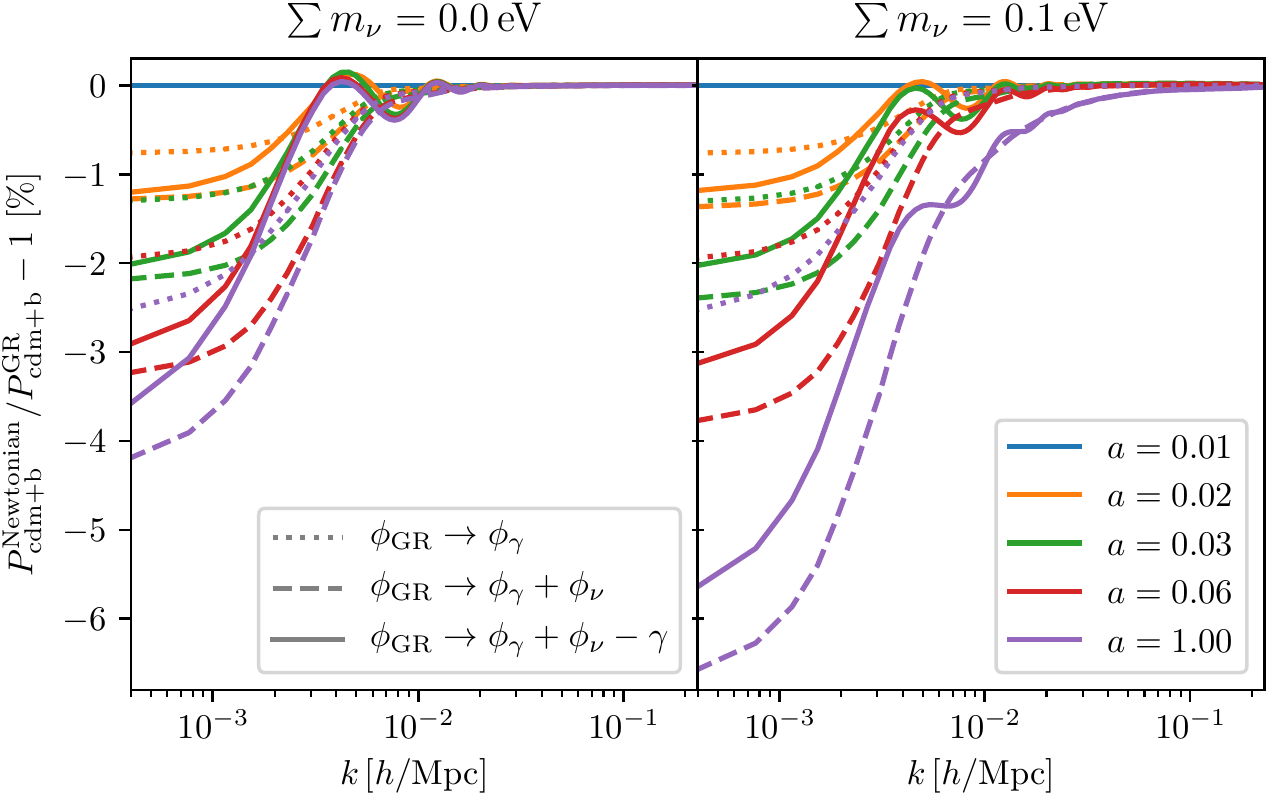}
\end{center}
\caption{Relative matter (CDM and baryons) power spectra with and without GR effects. Three levels of GR effects are considered; photon perturbations only (dotted lines), photon and neutrino perturbations (dashed lines) and photon, neutrino and metric perturbations (full lines). The left plot shows the case of massless neutrinos, the full lines of which are equivalent to Fig.~2 in \protect\cite{Brandbyge:2016raj}, the right plot shows the case of $\sum m_\nu=0.1\,\mathrm{eV}$. The power spectra are in $N$-body gauge.
\label{fig:relpower1}}
\end{figure}

In Fig.~\ref{fig:relpower1} we show ratios of matter power spectra with and without the corrections from $\phi_{\text{GR}}$ included. The effects from including the different terms ($\phi_\gamma$, $\phi_\nu$, $-\gamma$) in $\phi_{\text{GR}}$ are shown. It can be clearly seen that at early times models with and without neutrino mass behave identically because the neutrinos are still close to relativistic. Once neutrinos become non-relativistic the relative contribution from $\phi_\nu$ increases significantly and dominates over the other components, whereas the photons and the metric component ($\gamma$) are close to identical in the two cases. This is completely expected given the behaviour of $\phi_{\rm GR}$ seen in Fig.~\ref{fig:gamma}.

\begin{figure}[t]
\begin{center}
\includegraphics[width=0.8\textwidth]{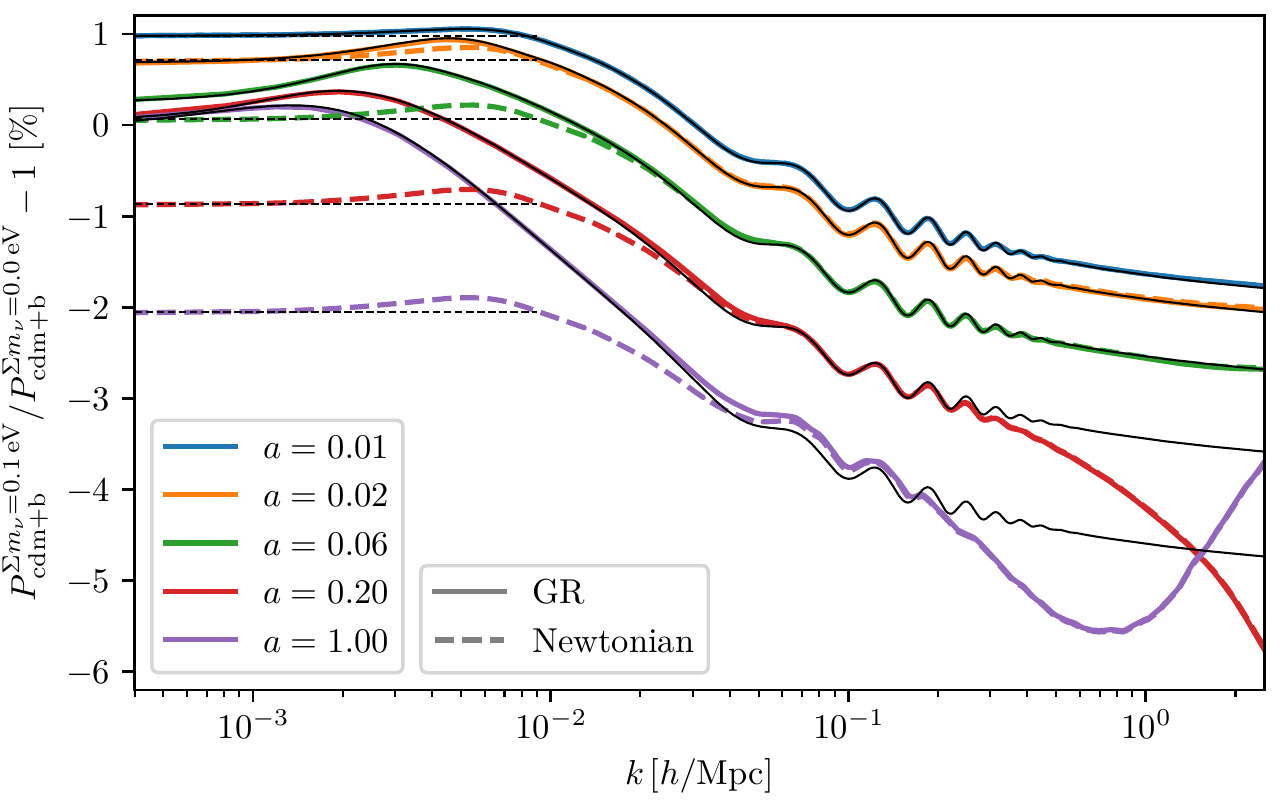}
\end{center}
\caption{Relative matter (CDM and baryons) power spectra between $\sum m_\nu=0.1\,\mathrm{eV}$ and $\sum m_\nu=0$ cosmologies. Dashed lines are without any GR effects and full lines with all GR effects included. Black lines indicate the corresponding linear results from \CLASS{}, where again full lines are in full GR (default \CLASS{}) and the dashed lines show the $k$-independent Newtonian growth rate. The power spectra are in $N$-body gauge.
\label{fig:relpower2}}
\end{figure}

In Fig.~\ref{fig:relpower2} we show the well-known suppression plot, comparing models with $\sum m_\nu=0.1\,\text{eV}$ to $\sum m_\nu=0$, with and without the full $\phi_{\rm GR}$ included. 

The dashed lines show results from running purely Newtonian simulations. 
We find the usual suppression in the semi-linear to non-linear regime (explained in detail in numerous other works, see e.g.\ \cite{Brandbyge:2008rv,Viel:2010bn,Agarwal:2010mt,Bird:2011rb,Villaescusa-Navarro:2013pva,Castorina:2015bma,Emberson:2016ecv,Adamek:2017uiq,AliHaimoud:2012vj,Liu:2017now,Brandbyge:2009ce,Banerjee:2016zaa,Bird:2018all}).

Notice that in the limit of small $k$ there are noticeable differences. At the initial time the model with non-zero neutrino mass has slightly more power, but over time the model with neutrino mass has slower growth of structure and therefore the power ratio drops with time. This phenomenon can be explained in the following way: At the initialisation time the amplitude of matter fluctuations $(\delta)$ is proportional to $\tau^2$, i.e.\ the conformal time at this particular $a$, squared. For the models shown here this $\tau^2$ differ by approximately $0.5\,\%$, and therefore the difference in power is approximately $1\,\%$.

Over time the Newtonian models lack any contribution from photon, neutrino and metric perturbations on large scales and since the matter density is lower in the model with neutrino mass, the matter fluctuations grow correspondingly slower, leading to suppression of power over time.

The thin, horizontal dashed lines show the ratio of solutions to the purely Newtonian linear perturbation equations for non-relativistic matter, i.e.\ the ratio of the growth functions, $D$, squared. Both models have almost the same background evolution. However, the model with massive neutrinos has no source term from the neutrinos acting on the CDM.
We normalise the ratio such that it matches exactly at the initial time. The fact that the simulations match the simple Newtonian linear theory result is a nice consistency check of the code.

The full curves show the result of simulations with $\phi_{\rm GR}$ included.
The thin black lines show the results from \CLASS{} (i.e.\ linear theory), and as can be seen the $N$-body results match exactly in the linear regime. As expected we see a slight increase in the ratio just before the non-linear scales (only clearly visible at $a=1$). For large $k$ we find the expected result, namely that there is an exact match between Newtonian and GR simulations. 

For the simulations with $\phi_{\rm GR}$ included the difference on large scales is far smaller. At the initialisation point the difference is the same as in the Newtonian case, since they start from the same \CLASS{} output. However, at later times the lack of cold dark matter is, to a large extent, compensated by the presence of neutrino and photon fluctuations. On super-horizon scales these are comparable in importance to the matter fluctuations and therefore the suppression becomes much less pronounced.

Finally, we note that the bump seen around $k \sim 6 \times 10^{-3} \, h/\text{Mpc}$ in the initial ratio arises from the difference in matter-radiation equality between the two models (see e.g.\ \cite{Lesgourgues:2006nd}), and that it propagates differently in the two models. In the Newtonian simulations it remains fixed in $k$-space, whereas in the GR case it moves to the left over time. This difference is caused by the GR corrections during evolution (i.e.\ it essentially amounts to the difference between the left and right panels in Fig.~\ref{fig:relpower1}).

\subsection{Comparison with PKDGRAV}

\begin{figure}[t]
\begin{center}
\includegraphics[width=0.8\textwidth]{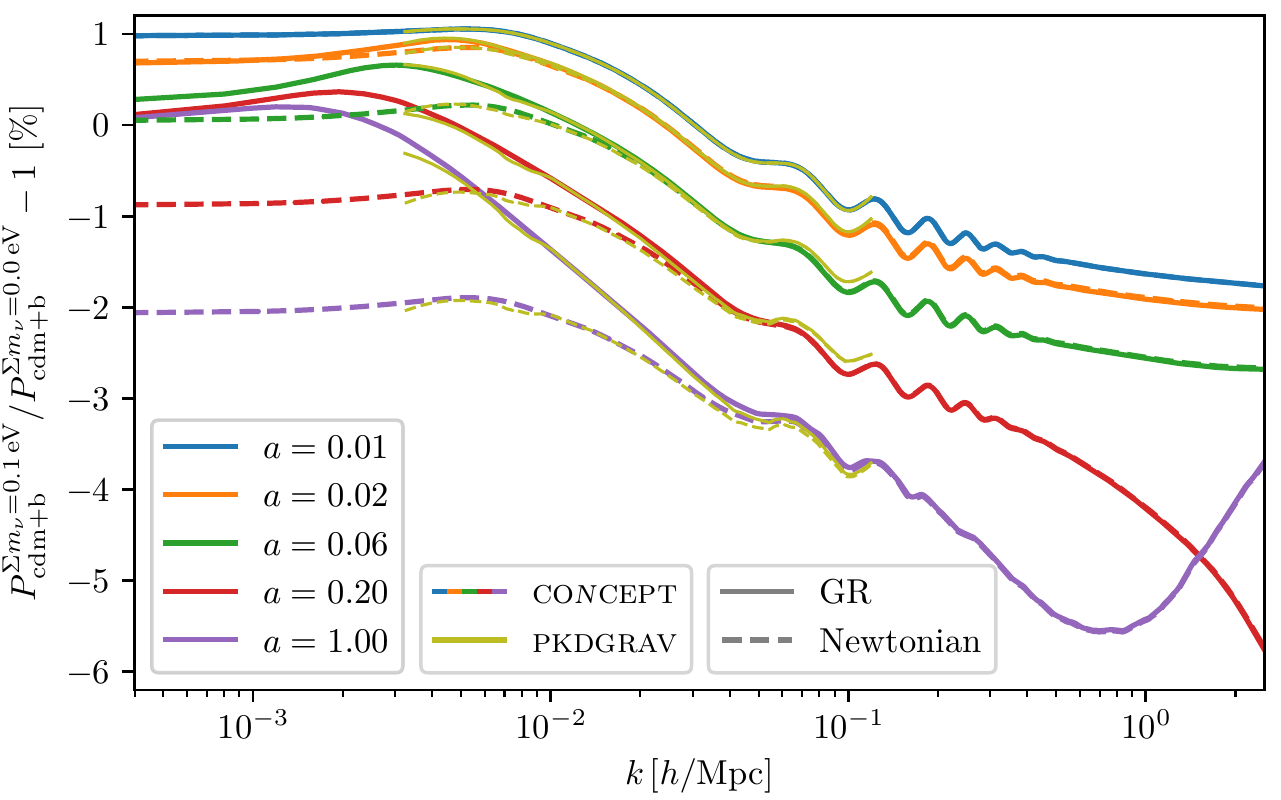}
\end{center}
\caption{Relative matter (CDM and baryons) power spectra between $\sum m_\nu=0.1\,\mathrm{eV}$ and $\sum m_\nu=0$ cosmologies, similar to Fig.~\ref{fig:relpower2}. The coloured lines show the \CONCEPT{} results and are identical to those in Fig.~\ref{fig:relpower2}. Cyan lines show the corresponding \PKDGRAV{} results.
\label{fig:relpower3}}
\end{figure}

In order to test the robustness of our calculation we have additionally implemented the GR effects in the state-of-the-art publicly available code \PKDGRAV{}.
Results from this exercise are shown in Fig.~\ref{fig:relpower3}.
As can be seen,  \PKDGRAV{} provides results which are identical to those of \CONCEPT{} to within a very small margin, even though the two codes are fundamentally different.

\PKDGRAV{} is a pure tree code, but with a grid structure implemented very recently precisely for the use case laid out in this paper. As seen in Fig.~\ref{fig:relpower3}, the results from the GR implementation in \PKDGRAV{} match those from the GR implementation in \CONCEPT{} very accurately. Through the newly added \CLASS{} mode of \PKDGRAV{}, all \PKDGRAV{} simulations use the exact same cosmology and initial conditions\footnote{Up to an effective change of random seed.} as the \CONCEPT{} simulations. Similarly, the box size is chosen as $(16384\,\text{Mpc}/h)^3$ and the simulations begin at $a=0.01$. The number of particles is however reduced to $512^3$, as we are only interested in the linear regime.

Fig.~\ref{fig:relpower3} do not show the \PKDGRAV{} lines at the lowest $k$ modes around $k\sim 10^{-3}\,h/\mathrm{Mpc}$, as here they begin to deviate from the expected results by a few percent. We can achieve agreement in this region by increasing the box size, which then simply moves the inaccurate region to the left. We suspect that this can be explained by the time stepping scheme used by \PKDGRAV{}, where all (major) time steps last for the same length of cosmic time. A new time stepping scheme based on the scale factor (at least at early times), rather than the cosmic time, is under construction.

Our \PKDGRAV{} simulations ended prematurely due to a hardware failure, and so the \PKDGRAV{} lines for $a = 1.00$ in Fig.~\ref{fig:relpower3} are really constructed from power spectra at $a = 0.50$, which we have extrapolated to $a = 1.00$ using linear theory.

\section{Discussion}

We have presented a framework for calculating the effect of light neutrinos, as well as photon and GR corrections in Newtonian $N$-body codes. The approach is based on the \COSIRA{} code presented in \cite{Brandbyge:2016raj}, in which radiation (photons and massless neutrinos) were included consistently to leading order in the $N$-body solver \GADGET{}. The method involves the realisation at all times of the radiation perturbation field and the scalar potential quantity $\gamma$ on a grid in the code. This grid is subsequently added to the ordinary potential grid to account for the effects of radiation and GR corrections to the Euler equation.

In this work we have extended the formalism to account for the possibility of massive neutrinos which complicates the calculation of $\gamma$ somewhat.
As in the case of massless neutrinos, we use the \CLASS{} code to calculate the quantities necessary to construct $\gamma$ in linear perturbation theory, i.e.\ the general relativistic potential correction $\gamma$ as well ass the energy density perturbations of photons and massive neutrinos. These are then realised on a grid in the $N$-body simulation.
We have tested the implementation in two different solvers: \CONCEPT{}, which is a PM code fully interfaced with \CLASS{}, and \PKDGRAV{}, which is a tree code to which has been added a potential grid in order to implement the effects.

We find that we can calculate the effects pertaining to the addition of massive neutrinos, photons and GR corrections at the per mille level on large scales, where structure formation is purely linear.
On smaller scales we find, as expected, that the effects of massive neutrinos are completely dominated by the absence of a clustering matter component, and that our results are identical to those found in a completely Newtonian $N$-body run.

The corrections studied here are typically at the level of a few percent on large scales, large enough that they should be included when comparing against data from future very large surveys such as EUCLID \cite{EUCLID} and LSST \cite{LSST}.

\section*{Acknowledgements}
This work was supported by the Villum Foundation. We thank Joachim Stadel for valuable comments on the draft. Jeppe Dakin thanks Joachim Stadel, Douglas Potter and Romain Teyssier for hospitality during a stay in which much of this work was carried out.

\appendix
\section{Computing $\gamma$}\label{appendix:gamma}
Here we will go through the steps necessary to calculate the quantity $\gamma$, appearing in \eqref{eq:gamma_from_perturbations}.
We will assume a spatially flat universe throughout this section. We start from the following definition of $\gamma$ from Eq.~4.12 in \cite{Fidler:2017pnb}:
\begin{align}
\gamma k^2  &\equiv -\left(\partial_\tau + \frac{\dot a}{a} \right) \dot H_\text{T} + 8 \pi G a^2 \Sigma ,\\
&= -\left(\partial_\tau + \frac{\dot a}{a} \right) \dot H_\text{T} + k^2 (\phi -\psi) \label{eq:gamma2} \,,
\end{align}
where the last line provides a convenient way of obtaining the total shear from quantities available in \CLASS{}.

Given that we run the simulations using $N$-body gauge we will now fix our discussion to this gauge.
In $N$-body gauge, we have $H^{\text{Nb}}_\text{T}=3\zeta$, where $\zeta$ is the comoving curvature perturbation, leading to
\begin{equation}
\dot H^{\text{Nb}}_\text{T} = 3\dot \zeta = 3 \frac{\dot a}{a} \left[\frac{\delta p^\text{com}}{\rho + p} + \sigma \right] \, ,
\end{equation}
where we have used the conservation equation for $\zeta$ in comoving gauge (Eq.~41 in \cite{Hu:2004xd}). 
The following gauge transformation of $\delta p^\text{com}$ is valid in both the Synchronous and Newtonian gauges since $B=0$ in those gauges:
\begin{equation}
\delta p^\text{com} = \delta p^\text{S/N} + \dot p \frac{\theta^\text{S/N}}{k^2} \,.
\end{equation}
Combining the equations, we find the following formula:
\begin{equation}
\dot H^{\text{Nb}}_\text{T} = 3 \frac{\dot a}{a} \frac{1}{\rho + p} \left[\delta p^\text{S/N} + \dot p \frac{\theta^\text{S/N}}{k^2}  + (\rho + p)\sigma \right] \, .
\end{equation}
Three of the quantities in this equation, $\delta p$, $\dot p$ and $\sigma$ are not readily available in the standard version of \CLASS{}. Thus, it is convenient to modify \CLASS{} slightly to output this quantity. We need a formula for $\dot p$ inside \CLASS{} that also includes non-cold dark matter. 
From Eq.~3.14 in \cite{Lesgourgues:2011rh} we find
\begin{equation}
\dot p_\alpha = -\frac{\dot a }{a} \left( 5p_\alpha - \mathfrak{p}_\alpha \right) \,,
\end{equation}
where $\mathfrak{p}$ is the pseudo-pressure defined in \cite{Lesgourgues:2011rh}. 
For any pressureless species, $\mathfrak{p}_\alpha \simeq p \simeq 0$, and for relativistic species we have $\mathfrak{p}_\alpha \simeq p$. We can then write the time-derivative of the total pressure in terms of the total pressure and $\mathfrak{p}_{\text{ncdm},\text{tot}}$:
\begin{equation}
\dot p = \sum_\alpha \dot p_\alpha =  -\frac{\dot a }{a} \left( 4 p +  p_{\text{ncdm},\text{tot}} - \mathfrak{p}_{\text{ncdm},\text{tot}} \right) \,.
\end{equation}

Using this prescription we have modified \CLASS{} to provide $\dot H^{\text{Nb}}_{\text T}$ in $N$-body gauge, which through \eqref{eq:gamma2} provides the quantity $\gamma$.


\bibliographystyle{utcaps}
\providecommand{\href}[2]{#2}\begingroup\raggedright
\endgroup

\end{document}